\documentstyle[prd,aps,epsf]{revtex}

\begin{document}


\newcommand{\beq}{\begin{equation}}
\newcommand{\eeq}{\end{equation}}
\newcommand{\bea}{\begin{eqnarray}}
\newcommand{\eea}{\end{eqnarray}}
\newcommand{\sla}{\!\!\!\!/ \,}
\newcommand{\slas}{\!\!\!\!\!/ \,}

\draft

\preprint{CERN-TH/99-351\\hep-ph/9911494}

\title{Four-Point Spectral Functions and Ward Identities in Hot QED }

\author{Hou Defu$^{1,2,3}$, M.E. Carrington$^{1,3}$, R. Kobes$^{1,4}$, and
        U. Heinz$^{5,}$\thanks{On leave from Institut f\"ur 
        Theoretische Physik, Universit\"at Regensburg, D-93040 
        Regensburg, Germany.}}
\address{${}^1$Winnipeg Institute for Theoretical Physics, Winnipeg, 
               Manitoba \\
${}^2$Institute of Particle Physics, Huazhong Normal University, 430070
      Wuhan, China \\
${}^3$Physics Department, Brandon University, Brandon, Manitoba, 
      R7A 6A9, Canada \\
${}^4$Physics Department, University of Winnipeg, Winnipeg, Manitoba, 
      R3B 2E9, Canada \\ 
${}^5$Theoretical Physics Division, CERN, CH-1211 Geneva 23, Switzerland}

\date{\today}
\maketitle

\begin{abstract}

We derive spectral representations for the different components of the 
4-point function at finite temperature in the real time formalism in 
terms of five real spectral densities. We explicitly calculate all these
functions in QED in the hard thermal loop approximation. The Ward 
identities obeyed by the 1-loop 3- and 4-point functions in real time 
and their spectral functions are derived. We compare our results with 
those derived previously in the imaginary-time formalism for retarded 
functions in hot QCD, and we discuss the generalization of our results 
to non-equilibrium situations.  

\end{abstract}



\section{Introduction}
\label{sec1}

One of the most significant advances in quantum field theory at finite
temperature\cite{LvW87,K89,bellac} is the effective perturbation theory 
of Braaten and Pisarski \cite{pisarski,mak} which is based on the 
resummation of hard thermal loops (HTLs). HTLs were originally 
obtained by computing one-loop diagrams in the imaginary time 
formulation of finite temperature field theory. HTLs are ultraviolet 
finite, gauge invariant and satisfy simple Ward identities 
\cite{hz,ft90,bp2,jac}. These  remarkable properties have been traced 
to the fact that HTLs describe simple semiclassical physics 
\cite{hz,Blz,kelly}. In particular, the HTLs can be derived from a set 
of classical kinetic equations \cite{hz,Blz} as well as from a non-local 
effective Lagrangian \cite{bp2}. The computation of HTLs is fairly
technical because of their non-trivial momentum and energy dependence, 
but can be simplified by using the Ward identities \cite{pisarski}.

The analytical structure of Green functions at zero or non-zero
temperature, in the imaginary or real-time formulations, is controlled
by their spectral densities, combined with asymptotic boundary
conditions. At finite temperature the spectral functions can also be
directly related to transport coefficients \cite{J93,WHZ96}. The study 
of spectral functions helps us to understand the quasi-particle structure 
of a field theory and to identify the microscopic processes underlying the
dynamics. But Green functions and their spectral functions are not
easy quantities to evaluate perturbatively at non-zero temperature,
especially for many-point functions. The two-point functions and their
spectral densities have been widely studied and applied to quark gluon
plasma (QGP) investigations \cite{kob,bellac,mak,J93}. Our knowledge
about many-point spectral densities is still much less comprehensive
\cite{kob1,evan1,ty,taylor,CH96,hh}. The spectral representations of
3-point functions for self-interacting scalar fields were discussed in
Refs.~\cite{kob1,evan1,CH96,hh}. In \cite{taylor} $3$- and $4$-point
spectral densities for pure gluon dynamics were calculated in the HTL
approximation using the imaginary time formalism (ITF) \cite{pisarski}, 
and in the same approximation the 3-point spectral density in QED has 
been calculated within the real time formalism (RTF) \cite{hh}. 

The popularity of the RTF is diminished by the technical complications 
resulting from the doubling of degrees of freedom \cite{chou}. However, 
in the ITF, one has to perform an analytic continuation of the imaginary 
external energy variables to the real axis, which is avoided in the RTF. 
In addition, the ITF is restricted to equilibrium situations, while the 
RTF can be extended to investigate non-equilibrium systems 
\cite{chou,keld,peterH}. 

In this paper we study 4-point functions in QED using the real-time 
formalism. We adopt the Keldysh or closed time path (CTP) contour
\cite{keld}. In Sec.~\ref{sec2} we introduce the Keldysh representation.
The spectral representations of the 4-point functions is derived 
in Sec.~\ref{sec3}. In Sec.~\ref{sec4} we use the Keldysh representation 
to calculate the 4-point HTL vertex functions in hot QED. The 4-point 
spectral densities in the HTL approximation are extracted in 
Sec.~\ref{sec5} and shown to degenerate to a single real spectral function.
In Sec.~\ref{sec6} we derive real-time Ward identities between the 3-
and 4-point HTLs and their spectral functions. The generalization of 
our results to certain types of non-equilibrium situations is discussed 
in Sec.~\ref{sec7}, while Sec.~\ref{sec8} gives a summary of our results.

\section{Keldysh representation}
\label{sec2}

We will use the Keldysh representation \cite{chou,keld} of the real-time 
formalism (RTF) throughout this paper. In this representation the 
single-particle propagator for free bosons has in momentum space the 
form \cite{chou,keld}
 \bea
   D(K)=\left (\begin{array}{cc} \frac{1}{K^2-m^2+i\epsilon} & 0\\
                                0 & \frac{-1}{K^2-m^2-i\epsilon}\\
               \end{array} \right )
   -2\pi i\, \delta (K^2{-}m^2) \left (\begin{array}{cc}
   n_B(k_0) & \theta ({-}k_0)+n_B(k_0)\\
   \theta (k_0)+n_B(k_0) & n_B(k_0) \\ \end{array} \right )
 \label{e1}
 \eea
where $K=(k_0,\bbox{k})$, $k=|\bbox{k}|$, $\theta $ denotes the step
function, and the equilibrium distribution function is given by 
$n_B(k_0)=1/[\exp(|k_0|/T)-1]$. For fermions the bare propagator can 
be written as
 \bea
   S(K)=(K \sla +m)\> \left [ \left (\begin{array}{cc}
   \frac{1}{K^2-m^2+i\epsilon} & 0\\
   0 & \frac{-1}{K^2-m^2-i\epsilon}\\
                          \end{array} \right )\right .
   +2\pi i\, \delta (K^2{-}m^2) 
      \left .\left (\begin{array}{cc}
      n_F(k_0) & -\theta ({-}k_0)+n_F(k_0)\\
      \theta (k_0)-n_F(k_0) & n_F(k_0) \\ 
                    \end{array} \right )\right ],
 \label{e2}
 \eea
where the Fermi-Dirac distribution is given by 
$n_F(k_0)=1/[\exp(|k_0|/T)+1]$. The components of these propagators 
are not independent, but fulfill the relation
 \beq
   G_{11}-G_{12}-G_{21}+G_{22}=0,
 \label{e3}
 \eeq
where $G$ stands for $D$ or $S$, respectively.

By an orthogonal transformation of these $2\times 2$ matrices one arrives
at the representation of the propagators in terms of advanced and retarded
propagators which was first introduced by Keldysh \cite{keld}. The three 
nonvanishing components of this representation are \cite{chou}
 \bea
   G_R & = & G_{11}-G_{12},
 \nonumber \\
   G_A & = & G_{11}-G_{21},
 \nonumber \\
   G_F & = & G_{11}+G_{22}.
 \label{e4}
 \eea
The inverted relations read
\bea
G_{11} & = & \frac{1}{2}\, (G_F+G_A+G_R),\nonumber \\
G_{12} & = & \frac{1}{2}\, (G_F+G_A-G_R),\nonumber \\
G_{21} & = & \frac{1}{2}\, (G_F-G_A+G_R),\nonumber \\
G_{22} & = & \frac{1}{2}\, (G_F-G_A-G_R).
\label{e5}
\eea
The relations (\ref{e3}--\ref{e5}) also hold for full propagators.

Using (\ref{e1}) and (\ref{e2}) in (\ref{e4}), the free propagators are
given in the Keldysh representation by
 \bea
   D_R(K) & = & \frac{1}{K^2-m^2+i\, \mbox{sgn}(k_0) \epsilon}\, ,
 \nonumber \\
   D_A(K) & = & \frac{1}{K^2-m^2-i\, \mbox{sgn}(k_0) \epsilon}\, ,
 \nonumber \\
   D_F(K)  & = & \bigl(1{+}2n_B(k_0)\bigr)\, \mbox{sgn}(k_0)\, 
   \Bigl(D_R(K)-D_A(K)\Bigr)
   = -2\pi i\, \bigl(1{+}2n_B(k_0)\bigr)\, \delta (K^2{-}m^2)
 \label{e6}
 \eea
for bosons and
 \bea
   S_R(K) & = & \frac{K\sla +m}{K^2-m^2+i\, \mbox{sgn}(k_0) \epsilon}
   =(K\sla +m)\bar D_R(K)\, ,
 \nonumber \\
   S_A(K) & = & \frac{K\sla +m}{K^2-m^2-i\, \mbox{sgn}(k_0) \epsilon}
   =(K\sla +m)\bar D_A(K)\, ,
 \nonumber \\
   S_F(K) & = & \bigl(1{-}2n_F(k_0)\bigr)\, \mbox{sgn}(k_0)\, 
   \Bigl(S_R(K)-S_A(K)\Bigr)
 \nonumber \\
   &=& -2\pi i\, (K\sla +m)\, \bigl(1{-}2n_F(k_0)\bigr)\, 
   \delta (K^2-m^2) = (K\sla +m)\bar D_F(K)
 \label{e7}
 \eea
for fermions, where the last equation in each of (\ref{e6}) and (\ref{e7}) 
is a consequence of the dissipation-fluctuation theorem. Although
$\bar D_{A,R} = D_{A,R}$ we keep the bar in our notation to more easily
recognize fermion and boson propagators in the calculations below.

\section{Spectral representation of the 4-point vertex}
\label{sec3}

In this section we briefly review some useful relations among the
different thermal components of the 4-point functions, and we derive 
their spectral representations. Some of these relations have already 
been reported in the literature \cite{kob,kob1,evan1,kob2} using different 
notation. We discuss these results at the end of this section. For  
any 1PI four-point function we define the four incoming external 
momenta as $P_1$, $P_2$,  $P_3$, and $P_4=-(P_1{+}P_2{+}P_3)$. In this 
section we suppress all indices except Keldysh indices.  

\subsection{Largest and smallest time equations}
\label{sec2b}

If $t_4$ is the largest time argument, one can obtain the ``largest time
equations'' \cite{kob,kob1,kob2}:

 \begin{mathletters}
 \label{9}
 \begin{eqnarray}
 \label{9a}
   \theta_{43}\,  \theta_{32}\,\theta_{21}\,(G_{abc1}+G_{abc2}) &=&
 \theta_{43}\,  \theta_{31}\,\theta_{12}\,(G_{abc1}+G_{abc2})=0\, ,
\\
   \label{9b}
 \theta_{42}\,  \theta_{23}\,\theta_{31}\,(G_{abc1}+G_{abc2})&=&
    \theta_{42}\,
\theta_{23}\,\theta_{13}\,(G_{abc1}+G_{abc2})=0\, ,
\\
 \label{9c}
\theta_{41}\,  \theta_{13}\,\theta_{32}\,(G_{abc1}+G_{abc2})&=&
 \theta_{41}\,  \theta_{12}\,\theta_{23}\,(G_{abc1}+G_{abc2})=0 \, ,
\\
 \label{9d}
 \theta_{32}\,  \theta_{21}\,\theta_{14}\,(G_{ab1c}+G_{ab2c}) &=&
    \theta_{31}\,  \theta_{12}\,\theta_{24}\,(G_{ab1c}+G_{ab2c})=0 \, ,
 \\
 \label{9e}
 \theta_{23}\,  \theta_{34}\,\theta_{41}\,(G_{a1bc}+G_{a2bc}) &=&
    \theta_{24}\,  \theta_{43}\,\theta_{31}\,(G_{a1bc}+G_{a2bc})=0 \, ,
 \\
 \label{9f}
   \theta_{13}\, \theta_{32}\,\theta_{24}\,(G_{1abc}+G_{2abc})& =&
   \theta_{12}\,\theta_{23}\, \theta_{34}\,(G_{1abc}+G_{2abc})=0 \, ,
 \end{eqnarray}
 \end{mathletters}
where $a,b, c$ can be either 1 or 2, and $\theta_{ij} \equiv \theta(t_i-t_j)$ 
is the step function.

By tilde conjugation (which replaces time ordering with anti-time ordering
\cite{chou,peterH}) one obtains equations of the form:
 \begin{mathletters}
 \label{10}
 \begin{eqnarray}
 \label{10a}
   \theta_{12}\,\theta_{23}\,\theta_{34}\,(\tilde G_{abc1}+\tilde G_{abc2})
   &=&
   \theta_{13}\,\theta_{32}\,\theta_{24}\,(\tilde G_{abc1}+\tilde G_{abc2})=0 
   \, ,
 \\
 \label{10b}
  \theta_{41}\,\theta_{12}\,\theta_{23}\,(\tilde G_{ab1c}+\tilde G_{ab2c}) 
  &=&
  \theta_{42}\,\theta_{21}\,\theta_{13}\,(\tilde G_{ab1c}+\tilde G_{ab2c})=0 
  \, ,
 \\
 \label{10c}
  \theta_{14}\,\theta_{43}\,\theta_{32}\,(\tilde G_{a1bc}+\tilde G_{a2bc}) 
  &=&
  \theta_{13}\,\theta_{34}\,\theta_{42}\,(\tilde G_{a1bc}+\tilde G_{a2bc})=0 
  \, ,
 \\
 \label{10d}
  \theta_{42}\,\theta_{23}\,\theta_{31}\,(\tilde G_{1abc}+\tilde G_{2abc}) 
  &=& 
  \theta_{43}\,\theta_{32}\,\theta_{21}\,(\tilde G_{1abc}+\tilde G_{2abc}) 
  =0 \, ,
 \end{eqnarray}
 \end{mathletters}
Eqs.~(\ref{9}) and (\ref{10}) are the analogues of the ``largest time
equations'' and ``smallest time equations'', respectively, of
Ref.~\cite{kob}. They will be used extensively in the derivation of the
spectral representations below. Their generalization to arbitrary 
$n$-point functions is straightforward.

\subsection{Derivation of spectral representations }
\label{sec3b}

One can construct the ``retarded--advanced'' vertex functions from 
the sixteen components of the real-time 4-point function as in 
\cite{CH96,chou}. The KMS conditions allows to reduce them
to the following seven combinations \cite{WH98,uh}:
\begin{mathletters}
 \label{11}
 \begin{eqnarray}
 \label{11a}
  G_{raaa} &=&
  G_{1111}+G_{1211}+G_{1121}+G_{1112}+G_{1122}+G_{1221}+G_{1212}+G_{1222},
\\
 \label{11b}
  G_{araa}&=&
  G_{1111}+G_{2111}+G_{1121}+G_{1112}+G_{1122}+G_{2121}+G_{2112}+G_{2122},
\\
 \label{11c}
  G_{aara}&=&
  G_{1111}+G_{2111}+G_{1211}+G_{1112}+G_{1212}+G_{2211}+G_{2112}+G_{2212},
\\
 \label{11d}
  G_{aaar}&=&
  G_{1111}+G_{2111}+G_{1211}+G_{1121}+G_{1221}+G_{2211}+G_{2121}+G_{2221},
\\
\label{11e}
  G_{rara}&=&
  G_{1111}+G_{1211}+G_{1112}+G_{1212}+G_{2121}+G_{2122}+G_{2221}+G_{2222},
\\
 \label{11f}
  G_{rraa}&=&
  G_{1111}+G_{1112}+G_{1121}+G_{1122}+G_{2211}+G_{2212}+G_{2221}+G_{2222},
\\
 \label{11g}
  G_{raar}&=&
  G_{1111}+G_{1121}+G_{1211}+G_{1221}+G_{2112}+G_{2122}+G_{2212}+G_{2222}.
 \end{eqnarray}
 \end{mathletters}
It is straightforward to show that the first four are the standard retarded 
vertex functions
 \begin{eqnarray}
  G_{raaa} = G_{R1}\,;~~~~G_{araa} = G_{R2}\,;~~~~G_{aara} = G_{R3}
  \,;~~~~G_{aaar} = G_{R4}.
 \end{eqnarray}
The last three we will call ``mixed retarded-advanced'' functions.

There are eight other vertices which one can obtain from (\ref{11}) 
using the KMS conditions \cite{WH98,uh}. We write down only the three 
which we need:
 \begin{mathletters}
 \label{11+}
 \begin{eqnarray}
 \label{11+a}
   G_{arar}&=&
   G_{1111}+G_{1121}+G_{2111}+G_{2121}+G_{1212}+G_{1222}+G_{2212}+G_{2222},
 \\
 \label{11+b}
   G_{arra}&=&
   G_{1111}+G_{1112}+G_{2111}+G_{2112}+G_{1221}+G_{1222}+G_{2221}+G_{2222},
 \\
 \label{11+c}
   G_{aarr}&=&
   G_{1111}+G_{1211}+G_{2111}+G_{2211}+G_{1122}+G_{1222}+G_{2122}+G_{2222}.
 \end{eqnarray}
 \end{mathletters}

We can re-express these vertices in terms of spectral functions by making 
use of various properties of theta functions. We will use 
 \beq
   \theta_{12} + \theta_{21} = 1,\quad
   \theta_{12} \theta_{21} = 0,\quad
   \theta_{23} \theta_{34} \theta_{24} = \theta_{23} \theta_{34}.
 \eeq

{\bf (i)} We begin with $G_{R1}$. We rewrite $G_{R1}$ as
\begin{equation}
G_{R1}=(\theta_{23}+\theta_{32})
(\theta_{34}+\theta_{43})(\theta_{24}+\theta_{42})G_{R1}.
\label{A1}
\end{equation}
Inserting the identities
 \begin{equation}
 \label{A2}
   \theta_{13} G_{R1} =\theta_{12} G_{R1} =
 \theta_{14} G_{R1} =G_{R1}
 \end{equation}
into (\ref{A1}) we obtain 
 \begin{eqnarray}
 \label{A4}
   G_{R1} &= &\theta_{12}\theta_{23} \theta_{34} G_{R1}+
              \theta_{12}\theta_{24} \theta_{43} G_{R1}+
              \theta_{14}\theta_{42} \theta_{23} G_{R1}
 \nonumber\\
  &+& \theta_{14}\theta_{43} \theta_{32} G_{R1}+
      \theta_{13}\theta_{34} \theta_{42} G_{R1}+
      \theta_{13}\theta_{32} \theta_{24} G_{R1}.
 \end{eqnarray}
Using the identities
 \begin{eqnarray}
 \label{A5}
    \theta(1ijk) G_{R2} =\theta_{1i} \theta_{ij}\theta_{jk} G_{R2}&=&0,
 \nonumber\\
    \theta(1ijk) G_{R3} =\theta_{1i} \theta_{ij}\theta_{jk} G_{R3}&=&0,
 \nonumber\\
    \theta(1ijk) G_{R4} =\theta_{1i} \theta_{ij}\theta_{jk} G_{R4}&=&0, 
    \quad (i,j,k=2,3,4)
 \end{eqnarray}
which result from conflicting $\theta$-functions, we have
 \begin{eqnarray}
 \label{A6}
   G_{R1} &= &[\theta(1234)+\theta(1243)+ \theta(1324)+\theta(1342)+
   \theta(1423)+\theta(1432)]
   \,(G_{R1}-G_{R2}+G_{R3}-G_{R4})
 \nonumber\\
   &=&[\theta(1234)+\theta(1243)+ \theta(1324)+\theta(1342)+
   \theta(1423)+\theta(1432)]\, \rho_{12}
 \end{eqnarray}
with
 \begin{eqnarray}
   \rho_{12}&=&G_{R1}-G_{R2}+G_{R3}-G_{R4}
 \label{A7}\\
   &=&G_{1211}-G_{2122}+2(G_{1212}-G_{2121})+G_{1112}-G_{2221}+G_{1222}
     -G_{2111}+G_{2212}-G_{1121}.
 \nonumber
 \end{eqnarray}

{\bf (ii)} For $G_{R2}, G_{R3}$ and $G_{R4}$ one proceeds similarly and
obtains
 \begin{mathletters}
 \label{A8}
 \begin{eqnarray}
 \label{A8a}
   G_{R2} &=&[\theta(2134)+\theta(2143)+ \theta(2314)+\theta(2341)+
              \theta(2413)+\theta(2431)](-\rho_{12}),
 \\
 \label{A8b}
   G_{R3} &= &[\theta(3124)+\theta(3142)+ \theta(3214)+\theta(3241)+
               \theta(3412)+\theta(3421)]\,\rho_{12},
 \\
 \label{A8c}
   G_{R4} &=&[\theta(4123)+\theta(4132)+ \theta(4213)+\theta(4231)+
              \theta(4312)+\theta(4321)](-\rho_{12}).
 \end{eqnarray}
 \end{mathletters}

{\bf (iii)} The procedure for the mixed retarded-advanced four-point 
functions is similar.  We write
\begin{equation}
G_{rraa}= (\theta_{12}+\theta_{21})(\theta_{14}+\theta_{41})
(\theta_{23}+\theta_{32})(\theta_{34}+\theta_{43})(\theta_{24}+\theta_{42})
G_{rraa}
\label{c1}
\end{equation}
and use 
\begin{eqnarray}
\theta(3ijk)G_{rraa}&=&0,\quad i,j,k=1,2,4, \nonumber
\\
\theta(4ijk)G_{rraa}&=&0,\quad i,j,k=1,3,4,
\end{eqnarray}
to obtain
\begin{eqnarray}
  G_{rraa}&=& 
  [\theta(1234)+\theta(1243)+\theta(1324)+\theta(1342)+\theta(1423)
  +\theta(1432)
 \nonumber \\
  &&\!\!+\theta(2134)+\theta(2143)+\theta(2314)+\theta(2341)+\theta(2413)+
  \theta(2431)]\, G_{rraa}\,.
\label{c2}
\end{eqnarray} 
Using the identities
\begin{eqnarray}
\theta(1ijk)\tilde G_{rraa}&=&0, \quad i,j,k=2,3,4,
\nonumber
\\
\theta(2ijk)\tilde G_{rraa}&=&0, \quad i,j,k=1,3,4
\end{eqnarray}
allows us to write 
\begin{eqnarray}
G_{rraa}=&& [\theta(1234)+\theta(1243)+\theta(1324)+\theta(1342)
            +\theta(1423)+\theta(1432)
 \nonumber \\
&&\!\!+\theta(2134)+\theta(2143)+\theta(2314)+\theta(2341)+\theta(2413)
+\theta(2431)]\,\rho_3
\label{c3}
\end{eqnarray}
with
\begin{equation}
 \label{c4}
  \rho_3=G_{rraa}-\tilde G_{rraa}\,.
\end{equation}
Analogously, we can derive
 \begin{mathletters}
 \begin{eqnarray}
   G_{rara}=&& [\theta(1234)+\theta(1243)+\theta(1324)+\theta(1342)
               +\theta(1423)+\theta(1432)
 \nonumber\\
   &&\!\!+\theta(3124)+\theta(3142)+\theta(3214)+\theta(3241)
         +\theta(3412)+\theta(3421)]\,\rho_4
 \label{c5a}
 \\
   G_{raar}=&&[\theta(1234)+\theta(1243)+\theta(1324)+\theta(1342)
              +\theta(1423)+\theta(1432)
 \nonumber\\
   &&\!\!+\theta(4123)+\theta(4132)+\theta(4213)+\theta(4231)
         +\theta(4321)+\theta(4312)]\,\rho_5
 \label{c5b}
 \end{eqnarray}
 \end{mathletters}
where
 \begin{mathletters}
 \label{c6}
 \begin{eqnarray}
 \label{c6a}
   \rho_4&=&G_{rara}-\tilde G_{rara}
 \\
 \label{c6b}
   \rho_5&=&G_{raar}-\tilde G_{raar} 
 \end{eqnarray}
 \end{mathletters}
Using the Fourier integral representation of the $\theta$ function,
 \begin{equation}
 \label{A20}
   \theta_{ij} = -\frac {1}{2\pi i}\int_{-\infty}^{\infty}
   d\Omega\, \frac{e^{-i\Omega(t_i-t_j)}}{\Omega+i\epsilon} \,,
 \end{equation}
it is straightforward to derive the spectral integral representations 
in momentum space:
 \begin{mathletters}
 \label{12}
 \begin{eqnarray}
 \label{12a}
   G_{R1}(\omega_1,\omega_2,\omega_3,\omega_4) &=& \oint 
   a_1^+[a_{12}^+(a_3^- + a_4^-) +a_{13}^+(a_2^- + a_4^-)+a_{14}^+( a_2^- +
   a_3^-)]\,\rho_{12}\,,
 \\
 \label{12b}
  G_{R2}(\omega_1,\omega_2,\omega_3,\omega_4) &=& \oint 
  a_2^+[a_{21}^+(a_3^- + a_4^-) +a_{23}^+(a_1^- + a_4^-)+a_{24}^+( a_1^- +
  a_3^-)](-\rho_{12})\,,
 \\
 \label{12c}
  G_{R3}(\omega_1,\omega_2,\omega_3,\omega_4) &=& \oint 
  a_3^+[a_{31}^+(a_2^- + a_4^-) +a_{32}^+(a_1^- + a_4^-)a_{34}^+( a_1^- +
  a_2^-)]\,\rho_{12}\,,
 \\
 \label{12d}
  G_{R4}(\omega_1,\omega_2,\omega_3,\omega_4) &=& \oint 
  a_4^+[a_{41}^+(a_2^- + a_3^-) +a_{42}^+(a_1^- + a_3^-)+a_{43}^+( a_1^- +
  a_2^-)](-\rho_{12})\,,
 \\
 \label{12e}
  G_{rraa}(\omega_1,\omega_2,\omega_3,\omega_4) &=& \oint 
  \Bigl(a_1^+[a_{12}^+(a_3^- + a_4^-) +a_{13}^+(a_2^- + a_4^-)
       +a_{14}^+( a_2^- + a_3^-)]
 \nonumber\\
  &&\ \,+a_2^+[a_{21}^+(a_3^- + a_4^-) +a_{23}^+(a_1^- + a_4^-)
        +a_{24}^+(a_1^- + a_3^-)]\Bigr)\,\rho_3\,,
 \\
 \label{12f}
  G_{rara}(\omega_1,\omega_2,\omega_3,\omega_4) &=& \oint 
  \Bigl(a_1^+[a_{12}^+(a_3^- + a_4^-) +a_{13}^+(a_2^- + a_4^-)
       +a_{14}^+( a_2^- + a_3^-)]
 \nonumber\\
  &&\ \, +a_3^+[a_{31}^+(a_2^- + a_4^-) +a_{32}^+(a_1^- + a_4^-)
         +a_{34}^+( a_1^- + a_2^-)]\Bigr)\,\rho_4
 \label{12g}
 \\
  G_{raar}(\omega_1,\omega_2,\omega_3,\omega_4) &=& \oint 
  \Bigl( a_1^+[a_{12}^+(a_3^- + a_4^-) +a_{13}^+(a_2^- + a_4^-)
        +a_{14}^+( a_2^- + a_3^-)]
 \nonumber\\
  &&\ \, +a_4^+[a_{41}^+(a_2^- + a_3^-) +a_{42}^+(a_1^- + a_3^-)
         +a_{43}^+( a_1^- + a_2^-)]\Bigr)\,\rho_5
 \end{eqnarray}
 \end{mathletters}
where
 \begin{mathletters}
 \label{13}
 \begin{eqnarray}
 \label{13a}
   \oint &=& \frac{i}{(2\pi)^3}\int d\Omega_1 d\Omega_2 d\Omega_3
 \\
 \label{13b}
   a_i^\pm &=& \frac{1}{\omega_i-\Omega_i\pm i\epsilon}
 \\
 \label{13c}
   a_{ij}^\pm &=& \frac{1}{\omega_i+\omega_i-\Omega_j-\Omega_j
                           \pm i\epsilon}\, ; 
 \quad i,j=1,2,3,4.
 \end{eqnarray}
 \end{mathletters}
The frequency arguments of the spectral functions under the integrals 
are $\rho_i(\Omega_1, \Omega_2, \Omega_3, \Omega_4)$ 
with $\Omega_1{+}\Omega_2{+}\Omega_3{+}\Omega_4{=}0$. The spatial 
momenta $\bbox{p}_1,\bbox{p}_2,\bbox{p}_3$, and 
$\bbox{p}_4{=}{-}(\bbox{p}_1{+}\bbox{p}_2{+}\bbox{p}_3)$ are the same 
on both sides of these equations and have therefore been suppressed. 

We have written the seven vertex functions (\ref{11}) in terms of one 
complex spectral density ($\rho_{12}\equiv \rho_1 + i \rho_2$) and three 
purely imaginary spectral densities $(\rho_3,\rho_4,\rho_5)$. This means 
that all sixteen 4-point functions can be expressed in terms of five 
real spectral densities:
 \begin{mathletters}
 \label{p}
 \begin{eqnarray}
 \label{pa}
  \rho_1&=&{\rm Re\,} [n_B^{-1}(p_{20})G_{2122} + 2n_B^{-1}(p_{20}
  + p_{40}) G_{2121} + n_B^{-1}(p_{40})G_{2221}
  - n_B^{-1}(p_{10})G_{1222} - n_B^{-1}(p_{30})G_{2212}]\,,
 \\ 
 \label{pb}
  \rho_2&=& - {\rm Im\,}[n_F^{-1}(p_{20})G_{2122} + 2n_F^{-1}(p_{20}
  + p_{40}) G_{2121} + n_F^{-1}(p_{40})G_{2221}
  - n_F^{-1}(p_{10})G_{1222} - n_F^{-1}(p_{30})G_{2212}]\,,
 \\
 \label{pc}
  \bar \rho_3 &=& -i \rho_3 = 2\,{\rm Im\,} G_{rraa}\,,
 \\
 \label{pd}
  \bar \rho_4 &=& -i \rho_4 = 2\,{\rm Im\,} G_{rara}\,,
 \\ 
 \label{pe}
  \bar \rho_5 &=& -i \rho_5 = 2\,{\rm Im\,} G_{raar}\,, 
 \end{eqnarray}
 \end{mathletters}
where we used the following relations in momentum space resulting from 
the KMS condition \cite{chou}:
 \begin{mathletters}
 \label{rp}
 \begin{eqnarray}
 \label{rpa}
  G_{1211}&=& e^{\beta p_{20}}\, G^*_{2122} \,, \quad 
  G_{1212} = e^{\beta (p_{20}+p_{40})}\, G^*_{2121}\,, \quad
  G_{1112} = e^{\beta p_{40}}\, G^*_{2221}\,,
 \\
 \label{rpb}
  G_{2111}&=& e^{\beta p_{10}}\, G^*_{1222}\,, \quad
  G_{1121} = e^{\beta p_{30}}\, G^*_{2212}\,, 
 \\
 \label{rpc}
  \tilde G_{rraa} &=& G^*_{rraa}\,, \quad
  \tilde G_{rara}=G^*_{rara}\,, \quad
  \tilde G_{raar}=G^*_{raar}\,.
 \end{eqnarray}
 \end{mathletters}
These results should be compared with the expressions derived in 
\cite{kob1} in which the spectral representations for the four retarded 
functions $G_{Ri}, ~\,i=1,2,3,4$ were given in terms of three real 
spectral densities. A relation between these three spectral densities 
exists which can be employed to reduce the number of independent spectral 
functions to two; this is consistent with our result. Spectral 
representations for the mixed retarded-advanced 4-point functions have 
also been previously discussed: in \cite{ty}, using a different definition 
of the mixed retarded-advanced 4-point functions, the spectral 
representation for one of these vertices was given in terms of 
six spectral densities.

We finally note that in deriving the spectral representations (\ref{12})
no use was made of the KMS condition. In this form, (with the spectral 
densities defined by (\ref{A4}), (\ref{c4}) and (\ref{c6})), 
Eqs.~(\ref{12}) remain valid out of thermal equilibrium.

\section{4-point  vertex functions in QED in the HTL approximation}
\label{sec4}

In this section, we calculate the seven 4-point vertex functions 
(\ref{11}) for QED in the HTL approximation. In the HTL approximation 
QED has two non-zero 4-point vertex functions, shown in Fig. [1] below: 
%
\begin{eqnarray}
\parbox{14cm}
{{
\begin{center}
\parbox{10cm}
{
\epsfxsize=8cm
\epsfysize=4cm
\epsfbox{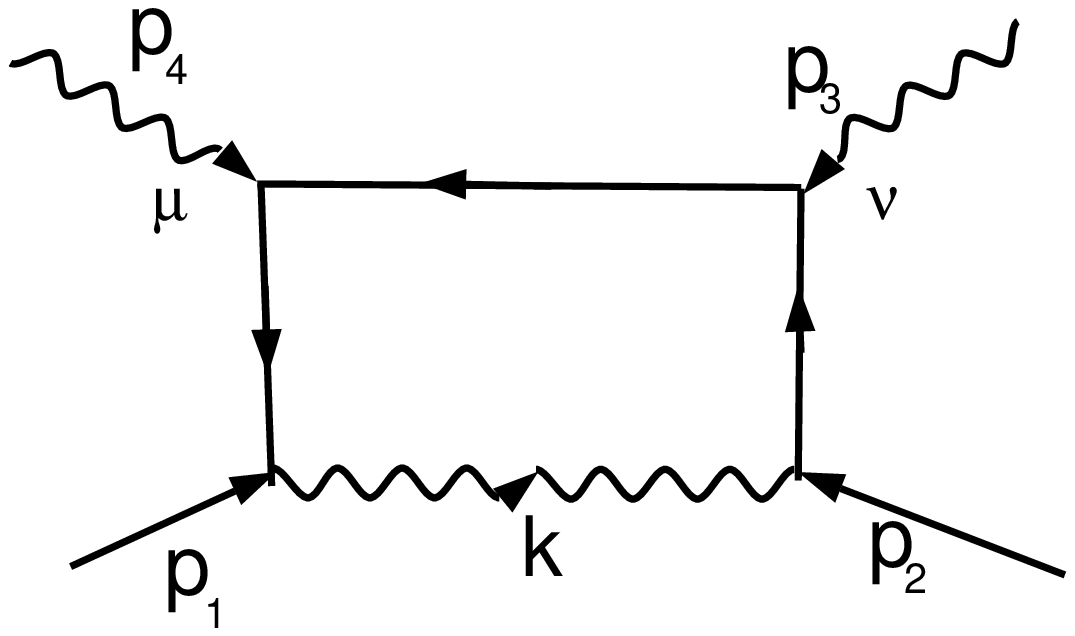}}
\end{center}
}}
\nonumber\\
\parbox{14cm}
{{
\begin{center}
\parbox{10cm}
{
\epsfxsize=8cm
\epsfysize=4cm
\epsfbox{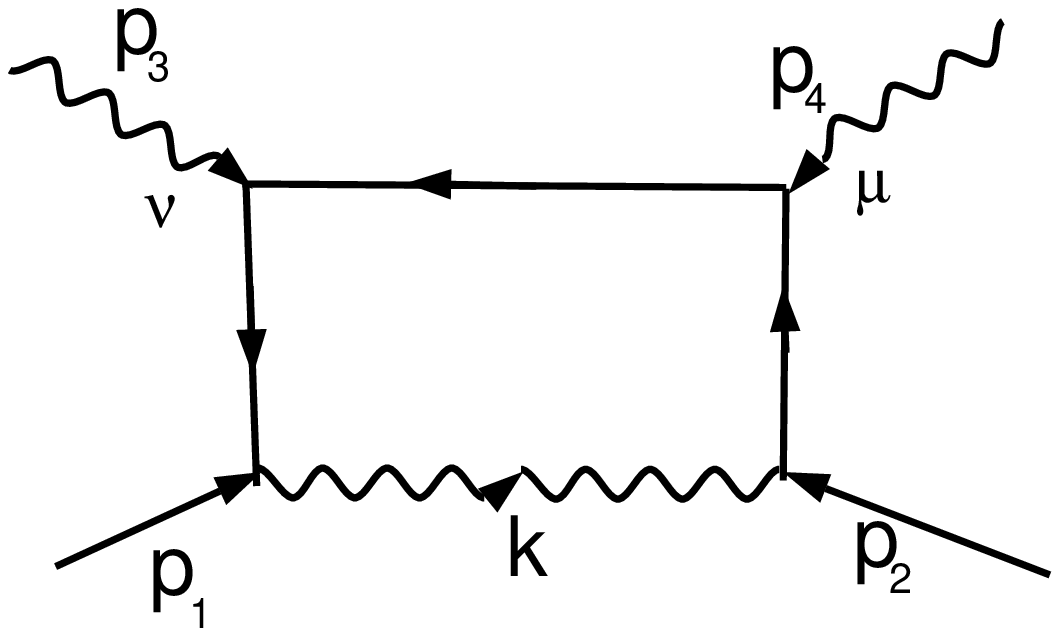}}\\
\parbox{14cm}{\small \center  Fig.~[1]: 4-point vertex in QED (a - upper graph)
and its cross term (b - lower graph).}
\label{F1}
\end{center}
}}
\nonumber
\end{eqnarray}
%
The other two 4-point vertex functions (four-photon vertex and four-fermion
vertex) are zero in the HTL approximation\cite{bellac}. We calculate 
the diagrams in Fig.~[1] as follows: the real-time Green functions are 
written in the Keldysh formalism using (\ref{e5}) and (\ref{11}). The 
photon propagator in the Feynman gauge is given by $-ig_{\mu\nu}D(K)$ 
where $D(K)$ is defined in (\ref{e6}). The fermion propagator is defined 
in (\ref{e7}). For the diagram in Fig.~[1a] we obtain
 \begin{mathletters}
 \label{A11}
\begin{eqnarray}
 \label{A11a}
  G_{R1}^{\mu\nu(a)}(P_1,P_2,P_3,P_4)
  &=&\frac{1}{2} \int \frac {d^4K}{(2\pi)^4} H^{\mu\nu}(K) 
  [a_1\bar a_2 \bar a_3\bar a_4 + r_1\bar r_2\bar r_3\bar r_4
  +f_1\bar a_2\bar a_3\bar a_4
 \nonumber\\
  &&\qquad\qquad\qquad
    +r_1\bar f_2\bar a_3\bar a_4+r_1\bar r_2\bar f_3\bar a_4
    +r_1\bar r_2\bar r_3\bar f_4]\,,
 \\
 \label{A11b}
  G_{R2}^{\mu\nu(a)}(P_1,P_2,P_3,P_4)
  &=&\frac{1}{2} \int \frac {d^4K}{(2\pi)^4} H^{\mu\nu}(K) 
  [a_1\bar a_2\bar a_3\bar a_4+r_1\bar r_2\bar r_3\bar r_4+
   f_1\bar r_2\bar r_3\bar r_4
 \nonumber\\
  &&\qquad\qquad\qquad
  +a_1\bar f_2\bar a_3\bar a_4
  +a_1r_2f_3a_4+a_1\bar r_2r_3 \bar f_4]\,,
 \\
 \label{A11c}
  G_{R3}^{\mu\nu(a)}(P_1,P_2,P_3,P_4)
  &=&\frac{1}{2} \int \frac {d^4K}{(2\pi)^4} H^{\mu\nu}(K) 
  [a_1\bar a_2\bar a_3\bar a_4+r_1\bar r_2\bar r_3\bar r_4+
   f_1\bar a_2\bar r_3\bar r_4
 \nonumber\\
  &&\qquad\qquad\qquad
  +r_1\bar f_2\bar r_3\bar r_4+a_1\bar a_2\bar f_3\bar a_4
  +a_1\bar a_2\bar r_3\bar f_4]\,,
 \\
 \label{A11d}
  G_{R4}^{\mu\nu(a)}(P_1,P_2,P_3,P_4)
  &=&\frac{1}{2} \int \frac {d^4K}{(2\pi)^4} H^{\mu\nu}(K) 
  [a_1\bar a_2\bar a_3\bar a_4+r_1\bar r_2\bar r_3\bar r_4+
   f_1\bar a_2\bar a_3\bar r_4
 \nonumber\\
  &&\qquad\qquad\qquad
  +r_1\bar f_2\bar a_3\bar r_4+r_1\bar r_2\bar f_3\bar r_4
  +a_1\bar a_2\bar a_3\bar f_4]\,,
 \\
 \label{A11e}
  G_{rara}^{\mu\nu(a)}(P_1,P_2,P_3,P_4)
  &=&\frac{1}{2} \int \frac {d^4K}{(2\pi)^4} H^{\mu\nu}(K) 
  [a_1\bar a_2\bar a_3\bar a_4+r_1\bar r_2\bar r_3\bar r_4+
   f_1\bar a_2\bar f_3\bar a_4+r_1\bar a_4 \bar f_2\bar f_3
 \nonumber\\
  &&\qquad\qquad\qquad
  +r_1\bar r_2\bar a_3\bar a_4+f_1\bar a_2\bar r_3\bar f_4
  +r_1\bar r_3\bar f_2\bar f_4+a_1\bar a_2 \bar r_3 \bar r_4]\,,
 \\
 \label{A11f}
  G_{rraa}^{\mu\nu(a)}(P_1,P_2,P_3,P_4)
  &=&\frac{1}{2} \int \frac {d^4K}{(2\pi)^4} H^{\mu\nu}(K) 
  [a_1\bar a_2\bar a_3\bar a_4+r_1\bar r_2\bar r_3\bar r_4+
   r_1\bar a_2\bar a_3\bar a_4+a_1\bar r_2\bar r_3\bar r_4
 \nonumber\\
  &&\qquad\qquad\qquad
  +f_1\bar f_2\bar a_3\bar a_4
  +f_1\bar f_3\bar r_2\bar a_4+f_1\bar f_4\bar r_2 \bar r_3]\,,
 \\
 \label{A11g}
  G_{raar}^{\mu\nu(a)}(P_1,P_2,P_3,P_4)
  &=&\frac{1}{2} \int \frac {d^4K}{(2\pi)^4} H^{\mu\nu}(K) 
  [a_1\bar a_2\bar a_3\bar a_4+r_1\bar r_2\bar r_3\bar r_4+
   a_1\bar a_2\bar a_3\bar r_4+r_1\bar r_2\bar r_3\bar a_4
 \nonumber\\
  &&\qquad\qquad\qquad
  +f_1\bar f_4\bar a_2\bar a_3
  +\bar f_2\bar f_4 r_1\bar a_3+\bar f_3\bar f_4 r_1\bar r_2 ]\,,
 \end{eqnarray}
 \end{mathletters}
where 
 \begin{equation}
 \label{Hmunu}
   H_{\mu\nu}(K) = e^4 \gamma_\alpha(K\sla - {P\sla}_1+m)
   \gamma_\mu({P\sla}_2 + {P\sla}_3 + K\sla +m)
   \gamma_\nu({P\sla}_2+ K\sla + m) \gamma_\alpha
 \end{equation}
and we used the notation $a_1=D_A(K)$, $r_1=D_R(K)$, 
$f_1= D_F(K)$ for photon propagators and $\bar a_j=\bar D_A(K_j)$,
$\bar r_j=\bar D_R(K_j)$, $\bar f_j=\bar D_F(K_j)$ [$j=2,3,4$ and 
$K_2{=}P_2{+}K$, $K_3{=}P_2{+}P_3{+}K$, $K_4{=}K{-}P_1$] for 
electron propagators. Note that the first two terms in each equation 
of (\ref{A11}) vanish by contour integration since all poles are 
located on the same side of the real axis. 

Before doing any momentum integrations, we can see immediately from
(\ref{A11}) that the one-loop retarded functions $(G_{Ri},~\, i=1,2,3,4)$ 
are linear in the distribution functions, since they are linear in 
the symmetric propagators $f_i$ and/or $\bar f_i$. Terms with higher 
powers of the distribution functions are explicitly cancelled in the 
Keldysh formalism. This type of cancellation also occurs for the 
2- and 3-point retarded-advanced functions \cite{baier}. In contrast, 
the mixed retarded-advanced functions are bilinear in the distribution 
functions. One of the advantages of the Keldysh formalism is that it 
allows us to immediately identify the leading order temperature 
dependence.  

In the HTL approximation, the external momenta are soft (of order 
$eT$ with $e\ll 1$) and the loop momenta are hard (of order $T$) 
\cite{pisarski,mak}. In this approximation, we can neglect the 
external momenta and the bare electron mass $m$ relative to the 
loop momenta. For $H_{\mu\nu}(K)$ one then obtains 
 \begin{equation}
 \label{HmunuHTL}
   H_{\mu\nu}(K)\approx 8e^4 K_\mu K_\nu K\sla \,.
 \end{equation}

We first consider $G_{R1}^a$. Fig.~[1a] gives
 \begin{eqnarray}
 \label{A12}
   G_{R1}^{\mu\nu(a)}(P_1,P_2,P_3,P_4) &=&
   \frac{1}{2} \int \frac {d^4K}{(2\pi)^4}\, H^{\mu\nu}(K) 
   [ f_1\bar a_2\bar a_3\bar a_4 + r_1\bar f_2\bar a_3\bar a_4 
   + r_1\bar r_2\bar f_3\bar a_4 + r_1\bar r_2 \bar r_3\bar f_4]
 \nonumber\\
   &=& I_1+I_2+I_3+I_4.
\end{eqnarray}
From (\ref{e6}), (\ref{e7}) we have
 \begin{eqnarray}
   a_i(K) &=& \bar a_i(K) = 
   \frac{1}{K^2-m_i^2-i{\rm sgn}(k_0)\epsilon}\, , 
 \nonumber\\
   r_i(K) &=& \bar r_i(K) = 
   \frac{1}{K^2-m_i^2+i{\rm sgn}(k_0)\epsilon}\, , 
 \nonumber\\
   f_1(K)&=&-2\pi i \delta(K^2)(1+2n_B(k))\, , 
 \nonumber\\
   \bar f_i(K) &=& -2\pi i \delta(K^2-m_i^2)(1-2n_F(k))\,,
 \label{raf}
 \end{eqnarray}
where $i{=}2,3,4$ and $m_2{=}m_3{=}m_4{=}m$ which stands for the 
electron mass. We calculate the contribution from each of the 
terms in (\ref{A12}) in the HTL approximation:  $K\sim T \gg P_1, 
P_2 , P_3, P_4, m$. For the temperature-dependent contributions we 
obtain
 \begin{mathletters}
 \label{A13}
 \begin{eqnarray}
 \label{A13a}
   I_1 &=& \frac{ie^4}{(2\pi)^3} \int dk\, k\, n_B(k) 
   \int d\Omega\,  {V^\mu V^\nu V\sla} \cdot b_{23}^- b_1^+ b_2^-\,,
 \\
 \label{A13b}
   I_2 &=& \frac{ie^4}{(2\pi)^3} \int dk\, k\, n_F(k) 
   \int d\Omega\, {V^\mu V^\nu V\sla}\cdot b_2^- b_3^- b_{12}^+\,,
 \\
 \label{A13c}
   I_3 &=& \frac{ie^4}{(2\pi)^3} \int dk\, k\, n_F(k) 
   \int d\Omega\, {V^\mu V^\nu V\sla} \cdot b_3^- b_4^- b_{23}^-\,,
 \\
 \label{A13d}
   I_4 &=& -\frac{ie^4}{(2\pi)^3} \int dk\, k\, n_F(k) 
   \int d\Omega\, {V^\mu V^\nu V\sla} \cdot b_1^+ b_4^- b_{12}^+\,,
 \end{eqnarray}
 \end{mathletters}
where
 \begin{mathletters}
 \label{13+}
 \begin{eqnarray}
 \label{13+a}
   b_i^\pm &=& \frac{1}{P_i \cdot V \pm i\epsilon}\,,
 \\
 \label{13+b}
   b_{ij}^\pm &=& \frac{1}{(P_i+P_j)\cdot V\pm i\epsilon}\,, \quad
   i,j=1,2,3,4\,,
 \end{eqnarray}
 \end{mathletters}
and $V=(1,\bbox{V})$ with $\bbox{V}= \bbox{k}/k$ a light-like
unit vector, and $\int d\Omega$ denotes integration over the 
orientation of $\bbox{V}$.

By interchanging $P_3$ and $P_4$ we obtain the contribution from Fig.~[1b]
to $G_{R1}$:
 \begin{equation}
 \label{A14}
   G_{R1}^{\mu\nu(b)} = J_1+J_2+J_3+J_4,
 \end{equation}
with
 \begin{mathletters}
 \label{15}
 \begin{eqnarray}
 \label{15a}
   J_1 &=& \frac{ie^4}{(2\pi)^3} \int dk\, k\, n_B(k) 
   \int d\Omega \, {V^\mu V^\nu V\sla} \cdot b_1^+ b_2^- b_{24}^-\,,
 \\
 \label{15b}
   J_2 &=& \frac{ie^4}{(2\pi)^3} \int dk\, k\, n_F(k) 
   \int d\Omega\, {V^\mu V^\nu V\sla} \cdot b_2^- b_4^- b_{12}^+\,,
 \\
 \label{15c}
   J_3 &=& \frac{ie^4}{(2\pi)^3} \int dk\, k\, n_F(k) 
   \int d\Omega\, {V^\mu V^\nu V\sla} \cdot b_3^- b_4^- b_{24}^-\,,
 \\
 \label{15d}
   J_4 &=& -\frac{ie^4}{(2\pi)^3} \int dk\, k\, n_F(k) 
   \int d\Omega\, {V^\mu V^\nu V\sla}\cdot b_1^+ b_4^- b_{12}^+\,.
 \end{eqnarray}
 \end{mathletters}
Combining (\ref{A13}) and (\ref{15}), the total contribution to 
$G^{\mu\nu}_{R1}$ from Fig.~[1] reads
 \begin{equation}
 \label{16}
   G_{R1}^{\mu\nu}(P_1,P_2,P_3,P_4) = 
   G_{R1}^{\mu\nu(a)}+G_{R1}^{\mu\nu(b)}
   = \frac{ie^2 m_\beta^2}{4\pi} \int d\Omega\, V^\mu V^\nu V\sla
     \cdot b_{23}^- b_{24}^-(b_1^+ - b_2^-)
 \end{equation}
where
 \beq
 \label{noneqm}
   m_\beta^2 = \frac{e^2}{2\pi^2} \int dk\, k\, (n_B(k)+n_F(k)) 
   = \frac{e^2T^2}{ 8}
 \eeq
is the usual expression for the square of the dynamically generated
thermal electron mass.

Proceeding similarly, we can calculate the other three retarded 4-point 
functions in finite temperature QED in the HTL approximation. The final 
results, including both diagrams in Fig.~[1], are
 \begin{mathletters}
 \label{17}
 \begin{eqnarray}
 \label{17a}
   G_{R2}^{\mu\nu}(P_1,P_2,P_3,P_4)
   &=& \frac{ie^2  m_\beta^2}{4\pi} \int d\Omega\, V^\mu V^\nu V\sla
   b_{23}^+ b_{24}^+ (b_1^- - b_2^+)\,,
 \\
 \label{17b}
   G_{R3}^{\mu\nu}(P_1,P_2,P_3,P_4)
   &=& \frac{ie^2m_\beta^2}{4\pi} \int d\Omega\,  V^\mu V^\nu V\sla
   b_{23}^+ b_{24}^+ (b_1^- - b_2^-)\,,
 \\
 \label{17c}
   G_{R4}^{\mu\nu}(P_1,P_2,P_3,P_4)
   &=& \frac{ie^2 m_\beta^2}{4\pi} \int d\Omega\,  V^\mu V^\nu V\sla
   b_{23}^- b_{24}^- (b_1^- - b_2^-)\,.
 \end{eqnarray}
 \end{mathletters}
Our results agree with those obtained in the ITF \cite{taylor}; to 
leading order, all of the retarded 4-point functions are proportional 
to $T^2$.

Next we consider the mixed retarded-advanced 4-point functions. We begin 
with $G^{\mu\nu}_{rara}$. The leading order temperature-dependent piece 
from Fig.~[1a] is 
 \bea
 \label{A11e1}
   G_{rara}^{\mu\nu(a)} (P_1,P_2,P_3,P_4)
   &=& \frac{1}{2} \int \frac {d^4K}{(2\pi)^4}\, H^{\mu\nu}(K) 
   [f_1\bar a_2\bar f_3\bar a_4 + r_1\bar f_2\bar f_3\bar a_4 +
    f_1\bar a_2\bar r_3\bar f_4 + r_1\bar r_3\bar f_2\bar f_4]
 \nonumber\\
   &=& T_1+T_2+T_3+T_4,
 \eea
where the functions $T_i\,;~~i=\{1,2,3,4\}$ represent the contributions of 
the four terms respectively. Inserting the propagators (\ref{raf}) 
and using the HTL approximation we find
 \begin{mathletters}
 \label{A15}
 \begin{eqnarray}
 \label{A15a}
   T_1 &=& \frac{e^4}{(2\pi)^3}c_\beta \int d\Omega\, 
   {V^\mu V^\nu V\sla}\, b_1^+ b_2^-(b_{23}^--b_{23}^+)\,,
 \\
 \label{A15b}
   T_2 &=& \frac{e^4}{(2\pi)^3}d_\beta \int d\Omega\,
   {V^\mu V^\nu V\sla}\, b_{12}^+ b_2^-(b_{3}^--b_{3}^+)\,,
 \\
 \label{A15c}
   T_3 &=& \frac{e^4}{(2\pi)^3}c_\beta \int d\Omega\,
  {V^\mu V^\nu V\sla}\, b_2^- b_{23}^+(b_1^+ - b_1^-)\,,
 \\
 \label{A15d}
   T_4 &=& \frac{e^4}{(2\pi)^3} d_\beta \int d\Omega\, 
   V^\mu V^\nu V\sla\, b_{3}^+ b_2^-(b_{12}^+ - b_{12}^-)\,,
 \end{eqnarray}
 \end{mathletters}
where
 \begin{eqnarray}
   c_\beta = && \int dk\, k\, [ n_B(k)(1{-}n_F(k)) - n_F(k)(1{+}n_B(k)) ]\,,
 \nonumber\\
   d_\beta = && -2 \int dk\, k\, n_F(k)(1{-}n_F(k))\,.
 \label{cd}
 \end{eqnarray}
Inserting the standard forms for the Bose-Einstein and Fermi-Dirac 
distributions we obtain in thermal equilibrium
 \begin{equation}
  c_\beta = 0\,,\quad  d_\beta = - (2 \ln 2)\, T^2\,.
 \label{cdeq}
 \end{equation}
The contribution from Fig.~[1b] to $G^{\mu\nu}_{rara}$ is
 \begin{equation}
 \label{rarab}
   G_{rara}^{\mu\nu(b)}=S_1+S_2+S_3+S_4
 \end{equation}
with
 \begin{mathletters}
 \label{A16}
 \begin{eqnarray}
 \label{A16a}
  S_1 &=& \frac{e^4}{(2\pi)^3}c_\beta \int d\Omega\, 
  {V^\mu V^\nu V\sla}\, b_1^+ b_2^-(b_{24}^--b_{24}^+)\,,
 \\
 \label{A16b}
   S_2 &=& \frac{e^4}{(2\pi)^3}d_\beta \int d\Omega\,
  {V^\mu V^\nu V\sla}\, b_{12}^+ b_2^-(b_{4}^--b_{4}^+)\,,
 \\
 \label{A16c}
   S_3 &=& \frac{e^4}{(2\pi)^3}c_\beta \int d\Omega\, 
   {V^\mu V^\nu V\sla}\, b_2^- b_{24}^+(b_1^+ - b_1^-)\,,
 \\
 \label{A16d}
   S_4 &=& \frac{e^4}{(2\pi)^3}d_\beta \int d\Omega\,
   {V^\mu V^\nu V\sla}\, b_{4}^+ b_2^-(b_{12}^+ - b_{12}^-)\,.
 \end{eqnarray}
 \end{mathletters}
Adding (\ref{A15}) and (\ref{A16}) we obtain
 \bea
   G_{rara}^{\mu\nu} = \frac{e^4}{(2\pi)^3} 
   \int d\Omega\, {V^\mu V^\nu V\sla} &&\Bigl[ d_\beta \,
   b_2^- (b_{12}^+(b_3^-+b_4^-)-b_{12}^-(b_3^++b_4^+))
   + c_\beta\,b_2^- (b_1^+(b_{23}^-+b_{24}^-) 
                   - b_1^-(b_{24}^++b_{23}^+))\Bigr]\,.
 \label{grara}
 \eea
In spite of (\ref{cdeq}) we keep the terms $\sim c_\beta$ because we 
will discuss the extension of these results to non-equilibrium 
situations in Sec.~\ref{sec7}.

The other two mixed retarded-advanced 4-point functions are similarly
calculated:
 \bea
   G_{raar}^{\mu\nu} = \frac{e^4}{(2\pi)^3} \int d\Omega\,
  {V^\mu V^\nu V\sla} && \Bigl[ d_\beta\, 
  b_1^+(b_{12}^+(b_3^-+b_4^-) - b_{12}^-(b_3^++b_4^+))
  + c_\beta\,b_2^- (b_1^+(b_{23}^-+b_{24}^-) 
                         -b_1^-(b_{24}^-+b_{23}^-))\Bigr]\,,
 \nonumber\\
   G_{rraa}^{\mu\nu} = \frac{e^4}{(2\pi)^3} \int d\Omega\,
  {V^\mu V^\nu V\sla} && \Bigl[c_\beta\, 
   (b_2^- b_1^+(b_{23}^-+b_{24}^-) - b_2^+ b_1^-(b_{24}^++b_{23}^+))\Bigr]\,.
 \label{mg}
 \eea
The last function vanishes in equilibrium since $c_\beta=0$. The other 
mixed retarded-advanced 4-point functions are again proportional to 
$T^2$. Furthermore, since $V^2{=}0$, all seven 4-point functions are 
traceless ($G^\mu_\mu=0$) in the HTL approximation.

\section{4-point spectral functions in hot QED in the HTL approximation}
\label{sec5}

From (\ref{16}), (\ref{17}), (\ref{grara}) and (\ref{mg}) one
obtains the following spectral representations for the 4-point 
retarded-advanced functions in the HTL approximation:
 \begin{mathletters}
 \label{18}
 \begin{eqnarray}
 \label{18a}
   G_{R1}^{\mu\nu}(\omega_1,\omega_2,\omega_3,\omega_4)&=&
   \frac{i}{(2\pi)^3}\int d\Omega_1 d\Omega_2 d\Omega_3 d\Omega_4
   \,\delta(\Omega_1{+}\Omega_2{+}\Omega_3{+}\Omega_4)
   \,\rho_{_{\rm HTL}}^{\mu\nu}(\Omega_1,\Omega_2,\Omega_3,\Omega_4)
   \, a_{23}^- a_{24}^+(a_1^+ - a_2^-)\,,
 \\
 \label{18b}
   G_{R2}^{\mu\nu}(\omega_1,\omega_2,\omega_3,\omega_4)&=&
   \frac{i}{(2\pi)^3}\int d\Omega_1 d\Omega_2 d\Omega_3 d\Omega_4
   \,\delta(\Omega_1{+}\Omega_2{+}\Omega_3{+}\Omega_4)
   \,\rho_{_{\rm HTL}}^{\mu\nu}(\Omega_1,\Omega_2,\Omega_3,\Omega_4)
   \, a_{23}^+ a_{24}^+(a_1^- - a_2^+)\,,
 \\
 \label{18c}
   G_{R3}^{\mu\nu}(\omega_1,\omega_2,\omega_3,\omega_4)&=&
   \frac{i}{(2\pi)^3}\int d\Omega_1 d\Omega_2 d\Omega_3 d\Omega_4
   \,\delta(\Omega_1{+}\Omega_2{+}\Omega_3{+}\Omega_4)
   \,\rho_{_{\rm HTL}}^{\mu\nu}(\Omega_1,\Omega_2,\Omega_3,\Omega_4)
   \, a_{23}^+ a_{24}^+(a_1^- - a_2^-)\,,
 \\
 \label{18d}
   G_{R4}^{\mu\nu}(\omega_1,\omega_2,\omega_3,\omega_4)&=&
   \frac{i}{(2\pi)^3}\int d\Omega_1 d\Omega_2 d\Omega_3 d\Omega_4
   \,\delta(\Omega_1{+}\Omega_2{+}\Omega_3{+}\Omega_4)
   \,\rho_{_{\rm HTL}}^{\mu\nu}(\Omega_1,\Omega_2,\Omega_3,\Omega_4)
   \,a_{23}^- a_{24}^-(a_1^- - a_2^-)\,,
 \\
 \label{18e}
   G_{rara}^{\mu\nu}(\omega_1,\omega_2,\omega_3,\omega_4)&=&
   \frac{i}{(2\pi)^3}\int d\Omega_1 d\Omega_2 d\Omega_3 d\Omega_4
   \,\delta(\Omega_1{+}\Omega_2{+}\Omega_3{+}\Omega_4)
   \,\rho_{_{\rm HTL}}^{\mu\nu}(\Omega_1,\Omega_2,\Omega_3,\Omega_4)
 \nonumber\\
   &&\cdot \Bigl[\alpha_1\, a_2^-(a_{12}^+(a_3^-+a_4^-)-a_{12}^-(a_3^++a_4^+))
   + \alpha_2\,(a_2^- a_1^+(a_{23}^-+a_{24}^-) - 
              a_2^- a_1^-(a_{24}^++a_{23}^+))\Bigr]\,,
 \\
 \label{18f}
   G_{raar}^{\mu\nu}(\omega_1,\omega_2,\omega_3,\omega_4)&=&
   \frac{i}{(2\pi)^3}\int d\Omega_1 d\Omega_2 d\Omega_3 d\Omega_4
   \,\delta(\Omega_1{+}\Omega_2{+}\Omega_3{+}\Omega_4)
   \,\rho_{_{\rm HTL}}^{\mu\nu}(\Omega_1,\Omega_2,\Omega_3,\Omega_4)
 \nonumber\\
   &&\cdot \Bigl[\alpha_1 a_1^+(a_{12}^+(a_3^-+a_4^-)-a_{12}^-(a_3^++a_4^+))
   + a_2 a_2^- a_1^+(a_{23}^-+a_{24}^-) - a_2^- a_1^-(a_{24}^-+a_{23}^-)
   \Bigr]\,,
 \\
 \label{18g}
   G_{rraa}^{\mu\nu}(\omega_1,\omega_2,\omega_3,\omega_4)&=&
   \frac{i}{(2\pi)^3}\int d\Omega_1 d\Omega_2 d\Omega_3 d\Omega_4
   \,\delta(\Omega_1{+}\Omega_2{+}\Omega_3{+}\Omega_4)
   \,\rho_{_{\rm HTL}}^{\mu\nu}(\Omega_1,\Omega_2,\Omega_3,\Omega_4)
 \nonumber\\
   &&\cdot
   \Bigl[\alpha_2\,(a_2^- a_1^+(a_{23}^-+a_{24}^-) - 
         a_2^+ a_1^-(a_{24}^++a_{23}^+))\Bigr]\,,
 \end{eqnarray}
 \end{mathletters}
where
 \begin{equation}
 \label{19}
   \rho_{_{\rm HTL}}^{\mu\nu}(\Omega_1,\Omega_2,\Omega_3,\Omega_4)
   = 2\pi^2 e^2  m_\beta^2 \int d\Omega\, V^\mu V^\nu V\sla
   \,\delta(\tilde P_1\cdot V)\, \delta(\tilde P_2\cdot V)
   \, \delta(\tilde P_3\cdot V)
 \end{equation}
(with $\tilde P^0_i \equiv \Omega_i,\, \tilde{\bbox{P}}_i = {\bbox{P}}_i;~$ 
$i{=}\{1,2,3\}$ on the r.h.s.) and
 \begin{equation}
   \alpha_1 = \frac{d_\beta e^4}{2i \pi^2 e^2 m_\beta^2}\,; 
 \quad
   \alpha_2 = \frac{c_\beta e^4}{2i \pi^2 e^2 m_\beta^2} \,.
 \label{const}
 \end{equation}
These equations also hold in the non-equilibium situations studied 
in Sec.~\ref{sec7}.  In equilibrium we use (\ref{cdeq}) to obtain
 \bea
   \alpha_1= i \frac{8\ln 2}{\pi^2}\,; \quad \alpha_2 = 0
 \label{const2}
 \eea
Note that the frequencies $\Omega_i$ should not be confused with the angular 
factor $\Omega$. From $V^2{=}0$ it follows that 
 \beq
 \label{rhomumu}
   (\rho_{_{\rm HTL}})^\mu_\mu=0\,.
 \eeq

In contrast to the general spectral representations (\ref{12}) of the 
4-point functions given in Sec.~\ref{sec3}, in the HTL approximation 
the spectral representations (\ref{18}) involve only a single, 
real-valued spectral density $\rho_{_{\rm HTL}}^{\mu\nu}$. This agrees 
with the result of Taylor \cite{taylor} who showed for QCD in the ITF 
that in the HTL approximation all spectral densities degenerate to a 
single function. Clearly, the spectral density (\ref{19}) is symmetric 
in the Lorentz indices and under interchange of the external momenta. 
Using the $\delta$-functions in (\ref{19}) it is also easy to show 
that $\rho_{_{\rm HTL}}^{\mu\nu}$ is transverse with respect to all 
external momenta:
 \begin{equation}
 \label{20}
   P^i_\mu\, \rho_{_{\rm HTL}}^{\mu\nu} = 0\, ,  \quad i=1,2,3,4.
 \end{equation}

A previous RTF calculation of the 3-point spectral function in hot QED 
in the HTL approximation gave \cite{uh}  
 \begin{equation}
 \label{20a}
   \rho_{_{\rm HTL}}^{\mu}(\Omega_1,\Omega_2,\Omega_3) =
   \frac {\pi e m_\beta^2}{2} \int d \Omega\, V^\mu V\sla\, 
   \delta(\tilde P_1\cdot V)\,\delta(\tilde P_2\cdot V)\,.
 \end{equation}
By comparing (\ref{19}) with (\ref{20}) we find the ``sum rule''
 \begin{equation}
 \label{20b}
   \int d\Omega_3\, \rho_{_{\rm HTL}}^{\mu 0}(\Omega_1,\Omega_2,\Omega_3,
   -(\Omega_1{+}\Omega_2{+}\Omega_3)) 
   = 4\pi e \, \rho_{_{\rm HTL}}^\mu(\Omega_1,\Omega_2,
   -(\Omega_1{+}\Omega_2))\,.
 \end{equation}
This agrees with the ITF results in \cite{taylor}. A similar sum rule 
exists between the 2- and 3-point spectral densities: the 2-point 
spectral density is given by
 \bea
   \rho_{_{\rm HTL}}(\Omega_1,-\Omega_1) = \frac{m_\beta^2}{2} 
   \int d\Omega\, V\sla\, \delta(\tilde P_1\cdot V)\,,
 \eea
and the temporal component of the 3-point spectral density 
$\rho_{_{\rm HTL}}^\mu (\Omega_1,\Omega_2, -(\Omega_1{+}\Omega_2))$ 
obeys the sum rule
 \bea
   \int d\Omega_2\, 
   \rho_{_{\rm HTL}}^0(\Omega_1,\Omega_2, -(\Omega_1{+}\Omega_2)) 
   = \pi\, e\, \rho_{_{\rm HTL}}(\Omega_1, -\Omega_1)\,.
\eea
We believe that analogous sum rules exist for the higher order 
$n$-point functions.

\section{Ward identities between 3- and 4-point functions in the 
         HTL approximation}
\label{sec6}

In this section, we verify the HTL Ward identities between the 3- and 
4-point vertex functions in QED, without doing momentum integrations. 
We split propagators in the integral expressions (\ref{A11}) by using
the identities \cite{mhm}
 \begin{equation}
   \bar D_{R/A} (K_2)\, \bar D_{R/A}(K_3)=\frac{1}{K_3^2-K_2^2}
   \Bigl(\bar D_{R/A}(K_2)-\bar D_{R/A}(K_3)\Bigr)
 \label{split}
 \end{equation}
This trick allows us to rewrite (\ref{A11}) as
 \begin{mathletters}
 \label{A11s}
 \begin{eqnarray}
 \label{A11sa}
   G_{R1}^{\mu\nu (a)} &=& e^4 \int \frac {d^4K}{(2\pi)^4} 
   \frac {4 K^\mu K^\nu K\sla}{2 P_3{\cdot}K}
   [f_1(\bar a_2-\bar a_3)\bar a_4 + r_1\bar f_2\bar a_4 -
    r_1 \bar f_3\bar a_4 + r_1(\bar r_2- \bar r_3)\bar f_4]\,,
 \\
 \label{A11sb}
    G_{R2}^{\mu\nu(a)} &=& e^4 \int \frac {d^4K}{(2\pi)^4} 
    \frac {4 K^\mu K^\nu K\sla}{2 P_3 \cdot K}
    [f_1(\bar r_2-\bar r_3)\bar r_4 + a_1\bar f_2\bar a_4 -
     a_1f_3a_4 + a_1(\bar r_2-\bar r_3) \bar f_4]\,,
 \\
 \label{A11sc}
    G_{R3}^{\mu\nu(a)} &=& e^4 \int \frac {d^4K}{(2\pi)^4} 
    \frac {4 K^\mu K^\nu K\sla}{2 P_3 \cdot K}
    [f_1(\bar a_2-\bar r_3)\bar r_4 + r_1\bar f_2\bar r_4 -
     a_1 \bar f_3\bar a_4 + a_1(\bar a_2-\bar r_3)\bar f_4]\,,
 \\
 \label{A11sd}
    G_{R4}^{\mu\nu(a)} &=& e^4 \int \frac {d^4K}{(2\pi)^4} 
    \frac {4 K^\mu K^\nu \not\!  K}{2 P_3 \cdot K}
    [f_1(\bar a_2-\bar a_3)\bar r_4 + r_1\bar f_2\bar r_4 - 
     r_1\bar f_3\bar r_4 + a_1(\bar a_2-\bar a_3) \bar f_4]\,,
 \\
 \label{A11Se}
    G_{rara}^{\mu\nu(a)} &=& e^4 \int \frac {d^4K}{(2\pi)^4} 
    \frac {4 K^\mu K^\nu K\sla}{2 P_3 \cdot K}
    [-f_1\bar f_3\bar a_4 + r_1(\bar r_2-\bar a_3)\bar a_4
     +f_1(\bar a_2-\bar r_3)\bar f_4
     +r_1\bar f_2\bar f_4 + a_1(\bar a_2-\bar r_3)\bar r_4]\,,
 \\
 \label{A11Sf}
    G_{rraa}^{\mu\nu(a)} &=& e^4 \int \frac {d^4K}{(2\pi)^4} 
    \frac {4 K^\mu K^\nu K\sla}{2 P_3 \cdot K}
    [r_1(\bar a_2-\bar a_3)\bar a_4 + a_1(\bar r_2-\bar r_3)\bar r_4
    +f_1\bar f_2\bar a_4
    -f_1\bar f_3\bar a_4 + f_1\bar f_4(\bar r_2 -\bar r_3)]\,,
 \\
 \label{A11Sg}
    G_{raar}^{\mu\nu(a)} &=& e^4 \int \frac {d^4K}{(2\pi)^4} 
    \frac {4 K^\mu K^\nu K\sla}{2 P_3 \cdot K}
    [a_1(\bar a_2-\bar a_3)\bar r_4 + r_1(\bar r_2 -\bar r_3)\bar a_4
    +f_1\bar f_4(\bar a_2-\bar a_3)
    +\bar f_2\bar f_4 r_1 - \bar f_3\bar f_4 r_1]\,,
 \end{eqnarray}
 \end{mathletters}
where we have used the HTL approximation to write $K_3^2-K_2^2 = 
2K\cdot P_3 + P_3^2 \approx 2K\cdot P_3$. 
%
\begin{eqnarray}
\parbox{14cm}
{{
\begin{center}
\parbox{10cm}
{
\epsfxsize=8cm
\epsfysize=6cm
\epsfbox{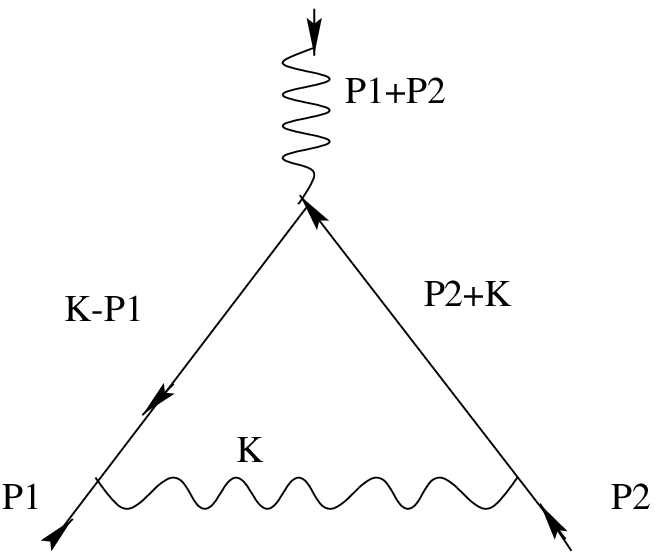}}\\
\parbox{14cm}{\small \center  Fig.~[2]: 3-point vertex in QED}
\label{F2}
\end{center}
}}
\nonumber
\end{eqnarray}
%
The integral expressions for the 3-point vertex functions in QED are 
obtained from Fig.~[2]. In the Feynman gauge we have
 \begin{mathletters}
 \label{A3p}
 \begin{eqnarray}
 \label{A3pa}
   G_{R}^\mu(P_1,P_2,P_3{+}P_4) &=& e^3 \int \frac {d^4K}{(2\pi)^4}
   (2K^\mu K\sla)[a_1\bar a_2 \bar a_4{+}r_1\bar r_2\bar r_4{+}f_1
    \bar r_2\bar r_4{+}a_1\bar f_2\bar a_4{+}a_1\bar r_2\bar f_4],
 \\
 \label{A3pb}
   G_{Ri}^\mu (P_1,P_2,P_3{+}P_4) &=& e^3 \int \frac {d^4K}{(2\pi)^4}
   (2K^\mu K\sla) [a_1\bar a_2\bar a_4{+}r_1\bar r_2\bar r_4{+}f_1
    \bar a_2\bar a_4{+}r_1\bar f_2\bar a_4{+}r_1r_2f_4],
 \\
 \label{A3pc}
   G_{Ro}^\mu(P_1,P_2,P_3{+}P_4) &=& e^3 \int \frac {d^4K}{(2\pi)^4}  
   (2K^\mu K\sla) [a_1\bar a_2\bar a_4{+}r_1\bar r_2\bar r_4{+}f_1
    \bar a_2\bar r_4{+}r_1\bar f_2\bar r_4{+}a_1\bar a_2\bar f_4],
 \\
 \label{A3pd}
   G_{F}^\mu(P_1,P_2,P_3{+}P_4) &=& e^3 \int \frac {d^4K}{(2\pi)^4} 
   (2K^\mu K\sla) [a_1\bar a_2\bar r_4{+}r_1\bar r_2\bar a_4{+}r_1
    \bar f_2\bar f_4{+}f_1\bar a_2\bar f_4],
 \\
 \label{A3pe}
   G_{Fi}^\mu(P_1,P_2,P_3{+}P_4) &=& e^3 \int \frac {d^4K}{(2\pi)^4} 
   (2K^\mu K\sla) [a_1\bar r_2\bar a_4{+}r_1\bar a_2\bar r_4{+}a_1
    \bar f_2\bar f_4{+}f_1\bar f_2\bar r_4],
 \\
 \label{A3pf}
   G_{Fo}^\mu(P_1,P_2,P_3{+}P_4) &=& e^3 \int \frac {d^4K}{(2\pi)^4} 
   (2K^\mu K\sla) [r_1\bar a_2\bar a_4{+}a_1\bar r_2\bar r_4{+}f_1
    \bar f_2\bar a_4{+}f_1\bar r_2\bar f_4],
 \end{eqnarray}
 \end{mathletters}
where for later convenience we have chosen to write on the l.h.s. 
$-(P_1{+}P_2)=P_3{+}P_4$. Comparing (\ref{A11s}) and (\ref{A3p}) we 
obtain the following Ward identities between the vertices shown in 
Fig.~[1a] and Fig.~[2]:
 \begin{mathletters}
 \label{wdA}
 \begin{eqnarray}
 \label{wdAa}
   P_{3\mu} G_{R1}^{\mu\nu(a)}(P_1,P_2,P_3,P_4) &=& 
   e \Bigl(G_{Ri}^\nu (P_1,P_2,P_3{+}P_4) - 
           G_{Ri}^\nu(P_1,P_2{+}P_3,P_4)\Bigr)\,,
 \\
 \label{wdAb}
   P_{3\mu} G_{R2}^{\mu\nu(a)}(P_1,P_2,P_3,P_4) &=&
   e \Bigl( G_R^\nu(P_1,P_2,P_3{+}P_4) -
            G_R^\nu(P_1,P_2{+}P_3,P_4)\Bigr)\,,
 \\
 \label{wdAc}
   P_{3\mu} G_{R3}^{\mu\nu(a)}(P_1,P_2,P_3,P_4) &=&
   e \Bigl( G_{Ro}^\nu (P_1,P_2,P_3{+}P_4) -
            G_{R}^\nu (P_1,P_2{+}P_3,P_4)\Bigr)\,,
 \\
 \label{wdAd}
   P_{3\mu} G_{R4}^{\mu\nu(a)}(P_1,P_2,P_3,P_4) &=&
   e \Bigl(G_{Ro}^\nu(P_1,P_2,P_3{+}P_4) 
          -G_{Ro}^\nu(P_1,P_2{+}P_3,P_4)\Bigr)\,,
 \\
 \label{wdAe}
   P_{3\mu} G_{rara}^{\mu\nu(a)}(P_1,P_2,P_3,P_4) &=&
   e \Bigl(G_{F}^\nu(P_1,P_2,P_3{+}P_4)-
           G_{Fo}^\nu(P_1,P_2{+}P_3,P_4)\Bigr)\,,
 \\
 \label{wdAf}
   P_{3\mu} G_{raar}^{\mu\nu(a)}(P_1,P_2,P_3,P_4) &=&
   e \Bigl(G_{F}^\nu(P_1,P_2,P_3{+}P_4)
          -G_{F}^\nu(P_1,P_2{+}P_3,P_4)\Bigr)\,,
 \\
 \label{wdAg}
   P_{3\mu} G_{rraa}^{\mu\nu(a)}(P_1,P_2,P_3,P_4) &=&
   e \Bigr(G_{Fo}^\nu(P_1,P_2,P_3{+}P_4)
          -G_{Fo}^\nu(P_1,P_2{+}P_3,P_4)\Bigr)\,.
 \end{eqnarray}
 \end{mathletters}
Similarly one can obtain the Ward identities satisfied by the diagrams 
in Fig.~[1b] and Fig.~[2]: 
 \begin{mathletters}
 \label{wdB}
 \begin{eqnarray}
 \label{wdBa}
   P_{3\mu} G_{R1}^{\mu\nu(b)}(P_1,P_2,P_3,P_4) &=&
   e \Bigl(G_{Ri}^\nu(P_1{+}P_3,P_2,P_4)
          -G_{Ri}^\nu(P_1,P_2,P_3{+}P_4)\Bigr)\,,
 \\
 \label{wdBb}
   P_{3\mu} G_{R2}^{\mu\nu(b)}(P_1,P_2,P_3,P_4) &=&
   e \Bigl(G_R^\nu(P_1{+}P_3,P_2,P_4)
          -G_R^\nu(P_1,P_2,P_3{+}P_4)\Bigr)\,,
 \\
 \label{wdBc}
   P_{3\mu} G_{R3}^{\mu\nu(b)}(P_1,P_2,P_3,P_4) &=&
   e \Bigl(G_{Ro}^\nu(P_1{+}P_3,P_2,P_4)
          -G_{Ro}^\nu(P_1,P_2,P_3{+}P_4)\Bigr)\,,
 \\
 \label{wdBd}
   P_{3\mu} G_{R4}^{\mu\nu(b)}(P_1,P_2,P_3,P_4) &=&
   e \Bigl(G_{Ri}^\nu(P_1{+}P_3,P_2,P_4)
          -G_{Ro}^\nu(P_1,P_2,P_3{+}P_4)\Bigr)\,,
 \\
 \label{wdBe}
   P_{3\mu} G_{rara}^{\mu\nu(b)}(P_1,P_2,P_3,P_4) &=&
   e \Bigl(G_{F}^\nu(P_1{+}P_3,P_2,P_4)
          -G_{F}^\nu(P_1,P_2,P_3{+}P_4)\Bigr)\,,
 \\
 \label{wdBf}
   P_{3\mu} G_{raar}^{\mu\nu(b)}(P_1,P_2,P_3,P_4) &=&
   - e G_{F}^\nu(P_1,P_2,P_3{+}P_4)\,,
 \\
 \label{wdBg}
   P_{3\mu} G_{rraa}^{\mu\nu(b)}(P_1,P_2,P_3,P_4) &=&
   e \Bigl(G_{Fo}^\nu(P_1{+}P_3,P_2,P_4)
          -G_{Fo}^\nu(P_1,P_2,P_3{+}P_4)\Bigr)\,.
 \end{eqnarray}
 \end{mathletters}
By combining (\ref{wdA}) and (\ref{wdB}) one obtains the following HTL 
Ward identities between the 3- and 4-point vertex functions in QED in 
the RTF:
 \begin{mathletters}
 \label{23}
 \begin{eqnarray}
 \label{23a}
   P_{3\mu} G_{R1}^{\mu\nu}(P_1,P_2,P_3,P_4) &=&
   e \Bigl(G_{Ri}^\nu(P_1{+}P_3,P_2,P_4)
          -G_{Ri}^\nu(P_1,P_2{+}P_3,P_4)\Bigr)\,,
 \\
 \label{23b}
   P_{3\mu} G_{R2}^{\mu\nu}(P_1,P_2,P_3,P_4) &=&
   e \Bigl(G_{R}^\nu(P_1{+}P_3,P_2,P_4)
          -G_{R}^\nu(P_1,P_2{+}P_3,P_4)\Bigr)\,,
 \\
 \label{23c}
   P_{3\mu} G_{R3}^{\mu\nu}(P_1,P_2,P_3,P_4) &=&
   e \Bigl(G_{R}^\nu(P_1{+}P_3,P_2,P_4)
          -G_{Ro}^\nu(P_1,P_2{+}P_3,P_4)\Bigr)\,,
 \\
 \label{23d}
   P_{3\mu} G_{R4}^{\mu\nu}(P_1,P_2,P_3,P_4) &=&
   e \Bigl(G_{Ro}^\nu(P_1{+}P_3, P_2,P_4)
          -G_{Ri}^\nu(P_1,P_2{+}P_3,P_4)\Bigr)\,,
 \\
 \label{23e}
   P_{3\mu} G_{rara}^{\mu\nu}(P_1,P_2,P_3,P_4) &=&
   e \Bigr(G_{F}^\nu(P_1{+}P_3,P_2,P_4)
          -G_{Fo}^\nu(P_1,P_2{+}P_3,P_4)\Bigr)\,,
 \\
 \label{23f}
   P_{3\mu} G_{raar}^{\mu\nu}(P_1,P_2,P_3,P_4) &=&
   -e G_{F}^\nu(P_1,P_2{+}P_3,P_4)\,,
 \\
 \label{23g}
   P_{3\mu} G_{rraa}^{\mu\nu}(P_1,P_2,P_3,P_4) &=&
   e \Bigl(G_{Fo}^\nu(P_1{+}P_3,P_2,P_4)
          -G_{Fo}^\nu(P_1,P_2{+}P_3,P_4)\Bigr)\,.
 \end{eqnarray}
 \end{mathletters}
Notice that in equilibrium the last equation gives simply $0=0$ (see
(\ref{cdeq}) and (\ref{mg})). These results are structurally similar 
to the zero temperature Ward identity \cite{Ward} and to the Ward 
identities between the retarded 3- and 4-point HTL vertices in QCD 
obtained within the imaginary time formalism \cite{bp2} or from 
kinetic theory \cite{Blz}. 

\section{Extension to Non-Equilibrium Situations}
\label{sec7}

The HTL method has been widely used in qualitative studies of the QGP
and in other explicit calculations at finite temperature \cite{mak}. 
This powerful method was derived within the ITF for equilibrium field
theory, and this is where its range of applicability can be clearly
defined \cite{pisarski,mak,bp2}: HTL resummation is required and 
applicable for the study of soft processes in a weakly interacting 
plasma, with momentum scale $p\sim g\,T \ll T$ ($g\ll 1$) much below 
the typical thermal or ``hard'' momenta in the medium. Realistic 
physical systems, on the other hand, are frequently out of equilibrium, 
which means that calculations must be carried out in the RTF. As long 
as the local momentum distribution $f(x,p)$ doesn't deviate too strongly 
from a thermal one, and can still be roughly characterized by a parameter 
$T$ which gives the scale for the typical momenta at point $x$, a similar
``hard loop'' (HL) resummation scheme should be applicable for the 
study of the dynamics of soft modes ($p \sim g\,T\ll T$) in a weakly 
interacting ($g\ll 1$) non-equilibrium system \cite{mhm}. This approach 
is expected to be reliable for equilibration processes requiring momentum 
exchanges which happen slowly, on a time scale $\tau\sim(g^2T)^{-1} 
\gg 1/p^0$. This reasoning does not apply to arbitrary non-equilibrium 
situations, but only to plasma states which are sufficiently close to 
global thermal equilibrium; on the other hand, deviations from chemical
equilibrium can be arbitrarily strong without invalidating the scheme.

In this section we discuss the generalization of our results to this 
particular type of non-equilibrium scenarios. Since we have used the 
Keldysh or closed time path contour, the procedure is straightforward.  
Equations (\ref{e1}) and (\ref{e2}) can be used in non-equilibrium 
situations by replacing the $x$-independent equilibrium Bose-Einstein 
and Fermi-Dirac distributions by non-equilibrium distributions
(Wigner functions) $f_B(x,p)$ and $f_F(x,p)$ which depend on the 
space-time coordinate and the four momenta \cite{chou}. Equations 
(\ref{e6}) and (\ref{e7}) remain valid for bare propagators, 
with the equilibrium distribution functions ($n_B, n_F$) replaced by 
the non-equilibrium ones ($f_B, f_F$), but the last equations in each 
of (\ref{e6}) and (\ref{e7}), (i.e. the fluctuation-dissipation theorem 
which reflects the KMS condition), no longer hold for full propagators 
\cite{mhm}. 

Equations (\ref{12}) give the spectral representations of the seven 
4-point functions (\ref{11}) in equilibrium. These expressions 
remain valid out of equilibrium since the KMS condition was not used
in their derivation. Out of equilibrium there are eight more 
4-point vertex functions ($G_{arrr}$, $G_{rarr}$, $G_{rrar}$, 
$G_{rrra}$, $G_{arar}$, $G_{aarr}$, $G_{arra}$ and $G_{rrrr}$) with
similar spectral representations. They can be derived using the same 
methods as those given in Sec.~\ref{sec3}.

The results in Secs.~\ref{sec4} and \ref{sec5} can be generalized to 
non-equilibrium situations by simply replacing in (\ref{noneqm}) and 
(\ref{cd}) the Bose-Einstein and Fermi-Dirac distributions 
$n_{B}(p^0)$ and $n_{F}(p^0)$ by non-equilibrium Wigner functions 
$f_{B}(x,p)$ and $f_{F}(x,p)$. The thermal mass $m_\beta$ in 
(\ref{noneqm}) and the factors $c_\beta,d_\beta$ in (\ref{cd}) become 
functions of $x$, $m(x),c(x),d(x)$, and it is only in the special case 
of thermal equilibrium that $c=0$ as obtained in (\ref{cdeq}). These 
generalized results correspond to the ``hard loop'' (HL) approximation 
in slightly off-equilibrium plasmas and are applicable in the 
situations discussed at the beginning of this section. We have shown 
that the 7 components of the 4-point function given in (\ref{11}) 
correspond to 5 spectral densities which degenerate in the HL 
approximation to a single real function. This result remains valid 
out of equilibrium, where the KMS conditions do not hold. The eight 
additional 4-point functions which must be included out of equilibrium 
will involve more spectral functions, but in the HL approximation it 
is straightforward to show that they degenerate to the same HTL 
spectral density (\ref{19}), with the thermal mass $m_\beta$ replaced 
by the generalized mass $m(x)$.  

The one-loop Ward identities in Sec.~\ref{sec6} were obtained without 
explicitly doing any momentum integrals; the result thus holds 
independently of the form of the distribution functions (which enter
through the functions $f_i$, $\bar f_i$ in (\ref{A11s}) and (\ref{A3p})).  
Out of equilibrium there will be additional Ward identities for the 
eight additional vertex functions which can be derived in the 
same way. For example, the Ward identities that correspond to the 
three functions $G_{aarr},~G_{arra},~ G_{arar}$ have been calculated, 
and they have the same form as those obtained in (\ref{23} e,f,g).

\section{Conclusions}
\label{sec8}

We have studied the 4-point vertex functions in the Keldysh formulation
of real-time thermal field theory \cite{keld}. This formalism has 
recently gained increased popularity because it avoids the need for 
analytical continuation associated with the imaginary time formalism, 
and it allows for a generalization to non-equilibrium situations. 
Starting from the largest time and smallest time equations which relate 
the 16 components of the 4-point real-time vertex function, we have 
derived spectral integral representations for the 7 retarded-advanced 
4-point functions (\ref{11}). We have explicitly calculated 
these 7 functions for hot QED at 1-loop order in the HTL approximation, 
using the Keldysh representation. In particular, we have demonstrated 
the usefulness of this representation for the calculation of many-point 
functions and showed how terms with high powers of the distribution 
functions cancel explicitly before any momentum integrations are done. 
The fully retarded vertex functions are found to be linear in the 
distribution functions \cite{baier}. In the HTL approximation, they 
have the same structure as their QCD counterparts which were derived 
in the ITF \cite{pisarski,taylor1} as well as from kinetic theory 
\cite{Blz}. We have also calculated the three mixed retarded-advanced 
4-point functions which are needed for a complete description of the 
real-time 4-point vertex and which have not previously appeared in 
the literature. They are quadratic in the distribution functions. One 
of them ($G_{rraa}$) was found to vanish in thermal equilibrium, while 
the other 4-point functions are all proportional to $T^2$.

We have derived spectral integral representations for the 7 
retarded-advanced 4-point functions (\ref{11}). The spectral densities 
were calculated explicitly for hot QED in the HTL approximation and found 
in this case to degenerate to a single real spectral function. Its trace 
over the Lorentz indices vanishes, and it is transverse with respect to 
all 4 external momenta. It obeys a simple sum rule with the correspondingly 
approximated 3-point spectral density.

By contracting the 7 retarded-advanced 4-point functions (\ref{11}) 
for QED in the HTL approximation with one of the photon momenta
we have derived one-loop Ward identities. This calculation was done 
without doing any momentum integrations, by comparing the integrands of 
the contracted HTL 4-point vertices with those of the HTL 3-point 
vertices. The resulting set of real-time Ward identities at finite 
temperature generalizes the zero temperature Ward identity and can 
be compared with the HTL Ward identity from the ITF and from kinetic 
theory. Due to the matrix form of the real-time thermal Green 
functions, the Ward identities have a more complex structure: one finds 
a whole class of finite temperature Ward identities which relate 
retarded-advanced 4-point functions to retarded-advanced 3-point 
vertex functions. If either the ingoing or the outgoing fermion leg 
has the largest time, the Ward identity involves only one kind of 
retarded-advanced 3-point vertex functions, and the right hand side 
vanishes when the contracted photon momentum ($P_3$) goes to zero. If 
one of the photon legs has the largest time, two different 
retarded-advanced 3-point functions are involved, and the right hand 
side no longer vanishes (Eqs.~(\ref{23}c-f); the corresponding component 
of the 4-point vertex is thus singular in the zero-momentum limit 
$P_3\to 0$ \cite{Boya}. Similar features were found for the Ward 
identities relating the HTL 3- and 2-point functions in hot QED 
\cite{hh}. 

Our Ward identities are not general, in the sense that they were
not directly derived by functional methods from the RTF path integral.
However, in the case of the 3-point functions it is known that the 
one-HTL Ward identities \cite{hh} are identical in structure 
with those satisfied by the exact vertex functions \cite{OTT}. We 
therefore expect the same to be true in the present case.

Finally, we have discussed the generalization of our results to 
certain classes of ``moderately non-equilibrium'' situations 
(Sec.~\ref{sec7}) in which a generalized HTL approximation (``hard 
loop approximation'') can be applied. Out of equilibrium the number 
of independent 4-point functions is larger, but in the HL approximation 
the spectral functions have the same structure as in equilibrium. All 
of the nice features of equilibrium HTLs persist in such ``moderately 
non-equilibrium'' situations.

\acknowledgments

We are grateful for interesting discussions with M. Thoma. This work
was supported by the National Natural Science Foundation of China (NSFC), 
the Deutsche Forschungsgemeinschaft (DFG), the Bundesministerium f\"ur 
Bildung und Forschung (BMBF), the Gesellschaft f\"ur 
Schwerionenforschung (GSI), and the Natural Sciences and Engineering 
and Research Council of Canada (NSERC). 


\end{document}